\documentclass[twoside,psfig]{article}

\catcode`\@=11
\long\def\@makefntext#1{
\protect\noindent \hbox to 3.2pt {\hskip-.9pt
$^{{\eightrm\@thefnmark}}$\hfil}#1\hfill}       

\def\@makefnmark{\hbox to 0pt{$^{\@thefnmark}$\hss}}    

\def\ps@myheadings{\let\@mkboth\@gobbletwo
\def\@oddhead{\hbox{}
\rightmark\hfil\eightrm\thepage}
\def\@oddfoot{}\def\@evenhead{\eightrm\thepage\hfil
\leftmark\hbox{}}\def\@evenfoot{}
\def\sectionmark##1{}\def\subsectionmark##1{}}

\oddsidemargin=\evensidemargin
\newcounter{sectionc}\newcounter{subsectionc}\newcounter{subsubsectionc}
\renewcommand{\section}[1] {\vspace{12pt}\addtocounter{sectionc}{1}
\setcounter{subsectionc}{0}\setcounter{subsubsectionc}{0}\noindent
    {\tenbf\thesectionc. #1}\par\vspace{5pt}}
\renewcommand{\subsection}[1] {\vspace{12pt}\addtocounter{subsectionc}{1}
    \setcounter{subsubsectionc}{0}\noindent
    {\bf\thesectionc.\thesubsectionc. {\kern1pt \bfit #1}}\par\vspace{5pt}}
\renewcommand{\subsubsection}[1] {\vspace{12pt}\addtocounter{subsubsectionc}{1}
    \noindent{\tenrm\thesectionc.\thesubsectionc.\thesubsubsectionc.
    {\kern1pt \tenit #1}}\par\vspace{5pt}}
\newcommand{\nonumsection}[1] {\vspace{12pt}\noindent{\tenbf #1}
    \par\vspace{5pt}}

\newcounter{appendixc}
\newcounter{subappendixc}[appendixc]
\newcounter{subsubappendixc}[subappendixc]
\renewcommand{\thesubappendixc}{\Alph{appendixc}.\arabic{subappendixc}}
\renewcommand{\thesubsubappendixc}
    {\Alph{appendixc}.\arabic{subappendixc}.\arabic{subsubappendixc}}

\renewcommand{\appendix}[1] {\vspace{12pt}
        \refstepcounter{appendixc}
        \setcounter{figure}{0}
        \setcounter{table}{0}
        \setcounter{lemma}{0}
        \setcounter{theorem}{0}
        \setcounter{corollary}{0}
        \setcounter{definition}{0}
        \setcounter{equation}{0}
        \renewcommand{\thefigure}{\Alph{appendixc}.\arabic{figure}}
        \renewcommand{\thetable}{\Alph{appendixc}.\arabic{table}}
        \renewcommand{\theappendixc}{\Alph{appendixc}}
        \renewcommand{\thelemma}{\Alph{appendixc}.\arabic{lemma}}
        \renewcommand{\thetheorem}{\Alph{appendixc}.\arabic{theorem}}
        \renewcommand{\thedefinition}{\Alph{appendixc}.\arabic{definition}}
        \renewcommand{\thecorollary}{\Alph{appendixc}.\arabic{corollary}}
        \renewcommand{\theequation}{\Alph{appendixc}.\arabic{equation}}
        \noindent{\tenbf Appendix \theappendixc #1}\par\vspace{5pt}}
\newcommand{\subappendix}[1] {\vspace{12pt}
        \refstepcounter{subappendixc}
        \noindent{\bf Appendix \thesubappendixc. {\kern1pt \bfit #1}}
    \par\vspace{5pt}}
\newcommand{\subsubappendix}[1] {\vspace{12pt}
        \refstepcounter{subsubappendixc}
        \noindent{\rm Appendix \thesubsubappendixc. {\kern1pt \tenit #1}}
    \par\vspace{5pt}}

\newcommand{\textlineskip}{\baselineskip=13pt}
\newcommand{\smalllineskip}{\baselineskip=10pt}

\def\eightcirc{
\begin{picture}(0,0)
\put(4.4,1.8){\circle{6.5}}
\end{picture}}
\def\eightcopyright{\eightcirc\kern2.7pt\hbox{\eightrm c}}

\newcommand{\copyrightheading}[1]
    {\vspace*{-2.5cm}\smalllineskip{\flushleft
    {\footnotesize Modern Physics Letters A, #1}\\
    {\footnotesize $\eightcopyright$\, World Scientific Publishing
     Company}\\
     }}

\newcommand{\publisher}[2]{{\begin{center}\footnotesize\smalllineskip
    Received #1\\
    Revised #2
    \end{center}
    }}

\renewenvironment{thebibliography}[1]
    {\frenchspacing
     \ninerm\baselineskip=11pt
     \begin{list}{\arabic{enumi}.}
        {\usecounter{enumi}\setlength{\parsep}{0pt}
     \setlength{\leftmargin 12.7pt}{\rightmargin 0pt} 
         \setlength{\itemsep}{0pt} \settowidth
    {\labelwidth}{#1.}\sloppy}}{\end{list}}

\newcounter{itemlistc}
\newcounter{romanlistc}
\newcounter{alphlistc}
\newcounter{arabiclistc}

\newcommand{\fcaption}[1]{
        \refstepcounter{figure}
        \setbox\@tempboxa = \hbox{\footnotesize Fig.~\thefigure. #1}
        \ifdim \wd\@tempboxa > 5in
           {\begin{center}
        \parbox{5in}{\footnotesize\smalllineskip Fig.~\thefigure. #1}
            \end{center}}
        \else
             {\begin{center}
             {\footnotesize Fig.~\thefigure. #1}
              \end{center}}
        \fi}

\newcommand{\tcaption}[1]{
        \refstepcounter{table}
        \setbox\@tempboxa = \hbox{\footnotesize Table~\thetable. #1}
        \ifdim \wd\@tempboxa > 5in
           {\begin{center}
        \parbox{5in}{\footnotesize\smalllineskip Table~\thetable. #1}
            \end{center}}
        \else
             {\begin{center}
             {\footnotesize Table~\thetable. #1}
              \end{center}}
        \fi}

\def\@citex[#1]#2{\if@filesw\immediate\write\@auxout
    {\string\citation{#2}}\fi
\def\@citea{}\@cite{\@for\@citeb:=#2\do
    {\@citea\def\@citea{,}\@ifundefined
    {b@\@citeb}{{\bf ?}\@warning
    {Citation `\@citeb' on page \thepage \space undefined}}
    {\csname b@\@citeb\endcsname}}}{#1}}

\newif\if@cghi
\def\cite{\@cghitrue\@ifnextchar [{\@tempswatrue
    \@citex}{\@tempswafalse\@citex[]}}
\def\citelow{\@cghifalse\@ifnextchar [{\@tempswatrue
    \@citex}{\@tempswafalse\@citex[]}}
\def\@cite#1#2{{$\null^{#1}$\if@tempswa\typeout
    {IJCGA warning: optional citation argument
    ignored: `#2'} \fi}}

\def\pmb#1{\setbox0=\hbox{#1}
    \kern-.025em\copy0\kern-\wd0
    \kern.05em\copy0\kern-\wd0
    \kern-.025em\raise.0433em\box0}

\def\fnt#1#2{\footnotetext{\kern-.3em
    {$^{\mbox{\scriptsize #1}}$}{#2}}}

\def\fpage#1{\begingroup
\voffset=.3in
\thispagestyle{empty}\begin{table}[b]\centerline{\footnotesize #1}
    \end{table}\endgroup}

\def\runninghead#1#2{\pagestyle{myheadings}
\markboth{{\protect\footnotesize\it{\quad #1}}\hfill}
{\hfill{\protect\footnotesize\it{#2\quad}}}}
\font\tenrm=cmr10
\font\tenit=cmti10
\font\tenbf=cmbx10
\font\bfit=cmbxti10 at 10pt
\font\ninerm=cmr9

\font\eightrm=cmr8

\def\qed{\hbox{${\vcenter{\vbox{            
   \hrule height 0.4pt\hbox{\vrule width 0.4pt height 6pt
   \kern5pt\vrule width 0.4pt}\hrule height 0.4pt}}}$}}

\newcommand{\gamfiv}{\gamma^5}
\newcommand{\fdual}{\widetilde{F}}

\begin{document}
\setlength{\textheight}{7.7truein}  

\runninghead{L.\ M.\ Rico,  M. Kirchbach}
{Causal Propagation of Spin-Cascades }

\normalsize\textlineskip
\thispagestyle{empty}
\setcounter{page}{1}

\copyrightheading{ }

\vspace*{0.88truein}

\fpage{1}
\centerline{\bf  Causal Propagation of Spin-Cascades
 }

\baselineskip=13pt
\vspace*{0.37truein}
\centerline{\footnotesize L.\ M.\ RICO, M.\ KIRCHBACH
\footnote{E-mail: mariana@ifisica.uaslp.mx}
}
\baselineskip=12pt
 \centerline{\footnotesize \it
Instituto de F\'{\i}sica, Universidad Aut.\ de San Luis Potos\'{\i},}
\baselineskip=10pt
\centerline{\footnotesize\it
Av. Manuel Nava 6, Zona Universitaria,}
\centerline{\footnotesize \it San Luis Potos\'{\i}, SLP 78290, M\'exico
}

\vspace*{10pt}

\publisher{(received date)}{(revised date)}

\vspace*{0.21truein}

\vspace*{10pt}

\textlineskip           
\vspace*{12pt}          

\begin{abstract}
We gauge the direct product of the Proca with the Dirac equation
that describes the coupling to the electromagnetic field of the
spin-cascade (1/2,3/2) residing in the four--vector spinor
$\psi_\mu$ and analyze propagation of its wave fronts in terms of
the Courant-Hilbert criteria. We show that the differential
equation under consideration is unconditionally hyperbolic and the
propagation of its wave fronts unconditionally causal. In this way
we proof that the irreducible spin-cascade embedded within
$\psi_\mu$  is free from the Velo-Zwanziger problem that plagues
the Rarita-Schwinger description of spin-3/2. The proof extends
also to the direct product of two Proca equations and implies
causal propagation of the spin-cascade (0,1,2) within an
electromagnetic environment.\\
\\
\emph{Keywords}: High-spins; causal propagation.

\end{abstract}

\vspace{3ex}

\section{Introduction}
The consistent description of high-spins within a covariant framework
is a long standing problem in particle physics.
The commonly used description  of fermions with spin higher than $1/2$
takes its origin from Refs.~\cite{PF}, \cite{RS}
which suggested to view fractional spin $J=(K+1/2)$ with $K$ integer
as the highest spin in
the totally symmetric rank--$K$ tensor spinor
$\psi_{\mu_1\mu_2...\mu_K}$ and describe it
by means of the Dirac equation,
\begin{equation}
(\not p  -m)\psi_{\mu_1\mu_2...\mu_K}=0\, ,
\label{RS_Dirac_tensor}
\end{equation}
as supplemented by the two
auxiliary conditions
\begin{eqnarray}
p ^{\mu_{1}}\psi_{\mu_1\mu_2...\mu_K}&=&0,\label{RS_Proca_tensor}\\
\gamma^{\mu_1}\psi_{\mu_1\mu_2...\mu_K}&=&0,
\label{RS_gamma_tensor}
\end{eqnarray}
with $p^\mu$ being the four momentum.
In so doing one restricts the degrees of freedom to $2(2J+1)$ which are then
associated with spin-$J$ particles and antiparticles.
The tensor-spinor representation spaces  reside in
the direct products of $K$ four-vector copies  with
the Dirac spinor $\psi=(1/2,0)\oplus (0,1/2)$ according to
\begin{eqnarray}
\psi_{\mu_1\mu_2...\mu_K}&= &\mbox{Sym}(1/2,1/2)_1\otimes
(1/2,1/2)_2\otimes ...\otimes (1/2,1/2)_{K} \nonumber \\&&\otimes
[(1/2,0)\oplus (0,1/2)]\, . \label{RSspinor}
\end{eqnarray}
They consist  in their rest frames of  $K$ parity doublets with
spins ranging from  $1/2^\pm $ to $(K-1/2)^\pm $ while the highest spin
$J=K+1/2$
remains a parity singlet:
\begin{equation}
\psi_{\mu_1\mu_2...\mu_K}\stackrel{\mbox{rest frame}}{\longrightarrow}
\frac{1}{2}^+, \frac{1}{2}^-;
\frac{3}{2}^+,\frac{3}{2}^-;...;
\left(K-\frac{1}{2}\right)^+, \left(K-\frac{1}{2}\right)^-;
\left(K+\frac{1}{2} \right)^\pi\, .
\label{Spin_Cascade}
\end{equation}
Here the parity of the highest spin is $(-1)^K$ for tensors,
and  $(-1)^{K+1}$ for pseudo-tensors.
Equation (\ref{Spin_Cascade}) illustrates the meaning of
$\psi_{\mu_1\mu_2...\mu_K}$ as spin-cascades.
The tensor-spinor of lowest rank is the
four-vector spinor, $\psi_\mu$. It is used in the description of spin-3/2,
in which case one encounters the shortest spin-cascade
\begin{equation}
\psi_\mu \stackrel{\mbox{rest frame}}{\longrightarrow}
\frac{1}{2}^+, \frac{1}{2}^-;\frac{3}{2}^-\, .
\label{Casc_3/2}
\end{equation}
Here we choose the polar four-vector spinor for concreteness.
Applied to $\psi_\mu$, Eqs.~(\ref{RS_Dirac_tensor})--(\ref{RS_gamma_tensor})
reduce the 16 degrees of freedom to 8 and associate them with particles
and anti-particles of spin-3/2 at rest. The first auxiliary condition in
Eq.~(\ref{RS_Proca_tensor}) excludes the spin-$0^+$ component of the
four-vector and thereby the spin-$1/2^+$ part of $\psi_\mu$, while the
second auxiliary condition in Eq.~(\ref{RS_gamma_tensor})
excludes its parity counterpart spin-$1/2^-$.
The  tensor-spinor framework can be given a Lagrangian formulation
upon establishing the most general form of a Lagrangian that leads to
the above three equations. Such a Lagrangian for
the four-vector--spinor $\psi_\mu$ can be found in \cite{MC} and reads
\begin{equation}
L(A)=\bar{\psi}^{\mu} \lbrack p_{\alpha}\Gamma_{\mu \ \, \nu}^{\
\,\alpha}-m g_{\mu\nu}\rbrack\psi^{\nu}\,,
\label{RS_Lgr}%
\end{equation}
where%
\begin{eqnarray}
p_{\alpha}\Gamma_{\mu\ \,\nu}^{\ \,\alpha}(A)\psi^{\nu}  & =&\not
p \psi_{\mu}+B(A)\gamma_{\mu}\not p
\,\gamma\cdot\psi+A(\gamma_{\mu}p\cdot
\psi+p_{\mu}\gamma\cdot\psi) +C(A)m\gamma_{\mu}\gamma\cdot \psi\,
\,,\label{Teil_1}\nonumber\\
A\neq\frac{1}{2},  && \,B(A)\equiv\frac{3}{2}A^{2}+A+\frac{1}{2},\
\, C(A)=3A^{2}+3A+1.
\label{Teil_3}%
\end{eqnarray}
The wave equation following from the above Lagrangian is obtained as
\begin{equation}
(\not p  -m)\psi_{\mu}+A\,(\gamma_{\mu}p\cdot\psi+p_{\mu}\,\gamma\cdot
\psi)+B(A)\,\left(  \,\gamma_{\mu}\not p  \,\gamma\cdot\psi\,\right)
+C(A)\,m\gamma_{\mu}\gamma\cdot\psi=0\,, \label{L(A)}%
\end{equation}
which for $A=-1$ can be written in a compact form as
\begin{equation}
(i\varepsilon_{\mu\nu\beta\alpha}\gamma^{5}\gamma^{\beta}p^{\alpha}%
-mg_{\mu \nu} +m\gamma_{\mu}\gamma_{\nu})\psi^{\nu}=0. \label{L(A=-1)}%
\end{equation}
Equations of this type  are equivalent to
\begin{eqnarray}
\frac{1}{2m}(\not p  + m)\psi_{\mu}  &  =& \psi_{\mu} \,,
\label{RS_Dirac}\\
(-g_{\mu\nu} + \frac{1}{m^{2}} p_{\mu } p_{\nu})\psi^{\nu}
&  =& -\psi_{\mu}\,,\label{RS_Proca}\\
\gamma^{\mu}\psi_{\mu}  &  =&0\,, \label{RS_What}%
\end{eqnarray}
known as the Rarita-Schwinger (RS) framework.\cite{RS}
Notice that for the sake of convenience of the point we are going to
make in the next section,
we here wrote the respective Dirac and Proca equations
(\ref{RS_Dirac}), and (\ref{RS_Proca}) in terms of covariant
projectors picking up spin-1/2$^{+}$ and spin-1$^{-}$ states,
respectively. Spin 3/2$^{+}$ needs an axial four vector.
Equation (\ref{RS_Dirac})--(\ref{RS_What})
suffer several inconsistencies one of them being
the non-causal propagation of the classical wave fronts of its solutions,
a result due to Ref.\cite{VZ1} and known as the Velo-Zwanziger problem.
Furthermore, the inverse of Eq.~(\ref{L(A)})  does not
relate to the spin 3/2 projector alone but is a more complicated
combination of various projectors \cite{Weda}.
This inconvenience can affect  the quantization procedure which is also
known to suffer inconsistencies \cite{SJ}.
\begin{quote}
We here made the case that the Velo-Zwanziger problem is not inherent
to the $\psi_\mu$ {\it representation space\/} by itself but rather to
its description within  the  Rarita-Schwinger framework.
To be specific, we shall show that the propagation of the spin cascade
$(1/2,3/2)$ when described in terms of the
direct product of the Proca and Dirac equations
does not suffer the Velo-Zwanziger problem but propagates strictly causally.
In addition, one gains coincidence between the nominator in
the  propagator (the latter being the inverse
of the wave equation) and  the projector  built from the states.
\end{quote}
Spin-cascade propagations are  of interest in the physics of
baryon resonances where according to Ref.~\cite{MK} (and references therein)
one observes well pronounced mass--
and parity degeneracies patterned after
the tensor-spinors of rank-1, 3, and 5, respectively,
but also possibly for the physics of the gravitino and the graviton, an
idea that has been put forward in  Refs.~\cite{KA},\cite{ADK}.
The idea of using spin-cascades as gauge fields in unified theories has been
pioneered by Kruglov and collaborators (see Ref.~\cite{SIK}
and references there in).

The paper is organized as follows. In the next section we briefly review
the essentials of the Velo-Zwanziger problem of
acuasal spin-3/2 propagation  in the light of the Currant-Hilbert
criteria.
Section III is devoted to the propagation properties of the
spin-cascades (1/2,3/2) and (0,1,2).
The paper ends with a brief summary.

\section{The Velo-Zwanziger problem of high-spin propagation}

The non-causal propagation of spin-3/2 within the gauged
Rarita-Schwinger framework has first been addressed in the work of
Giorgio Velo and Daniel Zwanziger\cite{VZ1}. {}For the sake of
self sufficiency of the presentation  we here highlight it in
brief. The main point of Ref.~\cite{VZ1} is that
Eq.~(\ref{L(A=-1)}) provided by the Lagrangian  (\ref{Teil_1}) is
not a genuine first order equation of motion because it does not
contain any time derivative of $\psi_{0}$ at all.

This defect shows up in the
(i) complete cancellation of all $\partial_{0}\psi_{0}$ terms in
Eq.~(\ref{L(A=-1)}) for any $\mu$,
(ii) complete cancellation of all the $\partial_{0}\psi_{\alpha}$ terms for
$\mu=0$, in which case one finds instead of a wave equation the constraint
\begin{equation}
\lbrack\mathbf{p}+(\mathbf{p}\cdot\,\mathbf{\gamma}-m)\mathbf{\gamma}]\cdot
\mathbf{\psi}=0\,, \label{prime_CS}%
\end{equation}
(iii) absence of $\psi_{0}$ in Eq.~(\ref{prime_CS}) that leaves the
time-component of the Rarita-Schwinger field undetermined.
The above deficits are caused by the constraints hidden in the
wave equation and could be tolerated only if remediable upon gauging.
Velo and Zwanziger gauge Eq.~(\ref{L(A=-1)}) in
Ref.~\cite{VZ1} in replacing $p_{\mu}$ by $\pi_{\mu}=p_{\mu}+eA_{\mu}$,
and succeed in constructing a genuine equation.
Their remedy procedure begins with first contracting the
gauged equation successively by $\gamma^{\mu}$ and $\pi^{\mu}$ and
obtaining the covariant gauged constraints as
\begin{eqnarray}
\gamma\cdot\psi &  =-&\frac{2}{3}\frac{ie}{m^{2}}\gamma^{5}\gamma
\cdot\widetilde{F}\cdot\psi\,,\label{1st_GSS}\\
\pi\cdot\psi &  =-& (\gamma\cdot\pi+\frac{3}{2}m)\frac{2}{3}\frac{ie}{m^{2}%
}\gamma^{5}\gamma\cdot{\widetilde{F}}\cdot\psi\,, \label{2nd_GSS}%
\end{eqnarray}
and ends with substituting Eqs.~(\ref{1st_GSS},\ref{2nd_GSS})
back into the gauged  Eq.~(\ref{L(A=-1)}). The
resulting new wave equation,
\begin{equation}
(\not \pi -m)\psi_{\mu}+(\pi_{\mu}+\frac{m}{2}\gamma_{\mu})\frac{2}{3}%
\frac{ie}{m^{2}}\gamma^{5}\gamma\cdot{\widetilde{F}}\cdot\psi\,=0\,,
\label{VZ_gauged}%
\end{equation}
is now a genuine one because it can be shown
to determine both $\psi_0$ and the time derivatives of $\psi_{\mu}$ for
any given $\mu$.

The final goal is to test hyperbolicity and causality
of Eq.~(\ref{VZ_gauged}) by means of the Courant-Hilbert criterion
\cite{CH} which requires the so called characteristic determinant of
the matrix containing the highest derivatives when
replaced by $n_{\mu}$, i.e. by the normals to the
characteristic surfaces, to vanish only for real $n_{0}$.
The Courant-Hilbert criterion is applied in fact not directly to
Eq.~(\ref{VZ_gauged}) but to its Hermitian form as obtained
by using  repeatedly Eqs.~(\ref{1st_GSS}) and (\ref{2nd_GSS}):
\begin{eqnarray}
\label{hermitian}
(\gamma\cdot\pi-m)\psi_\mu+(\pi_\mu&+&\frac{1}{2}m\gamma_\mu)
\frac{2ie}{3m^2}\gamfiv\gamma\cdot\fdual\cdot\psi\nonumber\\
&+&\frac{2ie}{3m^2}\fdual_\mu\cdot\gamma\gamfiv(\pi+\frac{1}{2}
m\gamma)\cdot\psi\nonumber\\
&+&\frac{2ie}{3m^2}\fdual_\mu\cdot\gamma
\gamma^5(\gamma\cdot\pi+2m)\frac{2ie}{3m^2}\gamma^5\gamma\cdot
\fdual\cdot\psi=0
\end{eqnarray}
The last equation represents now a system of partial differential equations
which would describe wave propagation phenomena provided it were
hyperbolic, something that
can be tested by exploiting the Courant-Hilbert criterion.
{}For this purpose it is sufficient to compute the normals $n_\mu$
to the characteristic surfaces. To find the
normals to the characteristic
surfaces passing through each point we replace
$i\partial_\mu$ by $n_\mu$ in the highest derivatives and
calculate the determinant $D(n)$ of the resulting coefficient
matrix, which is called the characteristic determinant. The
equation of motion Eq.~(\ref{hermitian}) will be hyperbolic if the
solutions $n^0$ to $D(n)=0$ are real for any $n_{\mu}=(n^0,{\mathbf{ n}})$.
By means of this prescription we are left with the following characteristic
determinant for the Rarita-Schwinger framework,
\begin{eqnarray}
D(n)&=&\left|\gamma\cdot n \
g_\mu^{\phantom{\mu}\nu}+\frac{2ie}{3m^2}n_\mu
\gamfiv\gamma\cdot\fdual^\nu+\frac{2ie}{3m^2}\fdual_\mu\cdot\gamma\gamfiv
n^\nu\nonumber \right.\\
& +&\left. \left(\frac{2ie}{3m^2}\right)^2\fdual_\mu\cdot\gamma
\gamfiv(\gamma\cdot n)\gamfiv\gamma\cdot\fdual^\nu \right|\, .
\end{eqnarray}
The covariant form of the latter is found to be,
\begin{equation}
\label{chareq}
 D(n)=(n^2)^4\left[n^2+\left(\frac{2e}{3m^2}\right)^2(\fdual\cdot
n)^2\right]^4=0\, .
\end{equation}
Equation~(\ref{chareq}) has the following four positive and
four negative roots
\begin{equation}
n_0=\pm \sqrt{{\mathbf{n}}^2}.
\end{equation}
Eight more  roots are found from
\begin{equation}
\label{roota} n^0=\pm \sqrt{ \frac{  ({\mathbf{n}}^2 - k^2
(\mathbf{B}\cdot \mathbf{n})^2)} {1-k^2{\mathbf{B}}^2}}\, =\pm
\frac{ \sqrt{ {\mathbf{n}}^2( 1-k^2{\mathbf{B}}^2\cos^2 \lambda
)(1-k^2{\mathbf{B}}^2)}} {1-k^2{\mathbf{B}}^2} \, ,
\end{equation}
respectively, where $\mathbf{B}$ stands for the magnetic field,
$k=\frac{2e}{3m^2}$, and $\lambda $ is the angle between $\mathbf{B}$
and $\mathbf{n}$.
Introducing the notion of the  "weak-field limit" to refer to the situation
in which there exists, for each space-time point, a Lorentz frame
such that the inequality,
\begin{equation}
\label{weak}
 k^2{\mathbf{B}}^2\leq 1\, ,
\end{equation}
is satisfied, we see that the roots given by
Eq.~(\ref{roota}) are real for any given
$n=(n^0,{\mathbf{n}})$, which establishes  hyperbolicity.

In the strong-field limit when the inequality (\ref{weak}) is no
longer satisfied, Eq.~(\ref{hermitian}) ceases to be hyperbolic
and is not suitable for the description of wave phenomena at all.

Moreover, in recalling that the maximum velocity of the signal
propagation  is the slope of the characteristic surfaces, one
immediately realizes that the characteristic surfaces determined
by Eq.~(\ref{chareq}) are not all tangent to the light cone.
Stated differently, one finds space-like characteristic surfaces
passing through all points where $F_{\mu\nu}$ is non-vanishing.
Consequently, such signals are propagated at velocities greater
than the speed of light. To see this it is just a matter of
realizing that there are time-like normals $n_\mu$ which satisfy
the characteristic equation (\ref{chareq}). Thus,
Eq.~(\ref{hermitian}) has characteristic surfaces which propagate
non-causally. In fact, for $n_\mu=(1,0,0,0)$  Eq.~(\ref{chareq})
takes the form of equality,
\begin{equation}
1=k^2{\mathbf{B}}^2\, ,
\end{equation}
and whenever $F_{\mu\nu}\neq 0$, there exists a Lorentz frame
where the latter equation holds valid.
The above considerations show that causal propagation
requires {\it unconditionally \/} hyperbolic wave equations.

\vspace{0.1cm} \noindent So far only few authors have proposed
solutions to the Velo-Zwanziger problem of the linear
Rarita-Schwinger Lagrangian \cite{RaSier}.\cite{Kruglov} In
Ref.~\cite{RaSier} Ra$\tilde n$ada and Sierra elaborate a method
which is equivalent to the  on-shell Rarita-Schwinger framework
but in their gauged formalism  the two auxiliary conditions have
been replaced by one differential equation. In effect, the number
of degrees of freedom increases from eight to twelve and the
theory in Ref.~\cite{RaSier} describes spin-1/2 and spin-3/2
particles of different masses. In applying the Courant-Hilbert
criterion to their wave equation, the authors of
Ref.~\cite{RaSier} prove that the wave fronts of the solutions of
their equation indeed do propagate causally. More recently,
Kruglov suggested two different wave equations, the first of which
is of second order and non-local while the second is local and
linear. \cite{Kruglov} None of the above equations suffers the
pathology of non-causal propagation of spin-3/2. The non-local
equation has twelve degrees of freedom of equal masses which are
associated with spin-3/2 and  spin-1/2. The local equation uses a
twenty dimensional space to embed  the four-vector spinor which
contains next to spin-3/2 also the two  copies of spin-1/2 of
opposite parities. Within the latter scenario the three different
spin sectors of $\psi_\mu$ appear characterized by three different
masses. Apparently, the mass-splittings between the spins is the
price to be paid for the  causal propagation of the solutions to
linear wave equations. Finally, single spins of the type
$(s,0)\oplus (0,s)$ have been shown to propagate always causally,
 results due to  Hurley \cite{Hurley_D4}, and
more recently to Ahluwalia and Ernst.\cite{AE}
On the one side, Hurley's  focus is  on the manifestly
hyperbolic nature of the generalized Feynman--Gell-Mann equations
for $(s,0)\oplus (0,s)$,
 \begin{equation}
(\pi ^{2}-m^{2})
\Psi^{(s,0)\oplus (0,s)}=
\frac{e}{2s} S^{\mu\nu}F_{\mu\nu}\Psi^{(s,0)\oplus (0,s)}\, ,
\label{Hurley}
\end{equation}
which are  of second order in the momenta,
and  obviously manifestly hyperbolic.
The  solutions of the generalized Feynman-Gell-Mann equations
therefore  propagate unconditionally causally.
Ahluwalia and Ernst construct the $(s,0)\oplus (0,s)$ propagation
from the different perspective of the representation space and  obtain it as
third order in the momenta. Their causality  proof is based on
the correct energy--momentum dispersion relations.
Despite their merits, the
 $(s,0)\oplus (0,s)$ representations are not as popular because
they are difficult to couple to the pion-nucleon or photon-nucleon system
due to dimensionality mismatch, a reason that still represents
an obstacle to their application in phenomenology.

\vspace{0.1cm}
\noindent
In the next section we (i) consider as a new  option a local but third order
single-mass wave equation describing the $(1/2,3/2)$ cascade residing in
$\psi_\mu$ and,
(ii) deliver the proof of the
causal propagation of its wave fronts within an electromagnetic environment.
The major appeal of the latter option is that it matches well
with the observed mass degeneracy of spin-1/2$^-$ and 3/2$^-$ baryons
such like, say, the  $S_{11}(1535)$ and $D_{13}(1520)$ resonances \cite{MK}
etc.
{}From that perspective it is desirable to have a spin-cascade equation
that is characterized by a single  mass.

\section{Cascade--spin description and causal propagation}

In order to find the wave equation for the spin-cascade of
interest it is quite instructive to go back to
Eq.~(\ref{RSspinor}) and to recall the wave equation for the
four-vector spinor. Although one encounters various equations for
the $(1/2,1/2)$ representation space in the literature,
\cite{Ryder},\cite{hh}, \cite{Mauro},\cite{VZ2} all they reflect
different facets of  the following equation and related auxiliary
condition:
\begin{eqnarray}
\left[(p^2-m^2)[g]^{\nu}_{\ \, \mu}-p_\mu
p^\nu\right]A^\mu&=&0, \label{Proca_main}\\\
p_\mu A^\mu&=&0\, .
\label{Proca_aux}
\end{eqnarray}
The latter equations imply that the four degrees of freedom
of the $(0^+,1^-)$ spin cascade
in $(1/2,1/2)$ have been reduced to three and are
associated with spin-1 at rest. Now we consider the
direct product of Eq.~(\ref{Proca_aux}) with the Dirac equation
\begin{eqnarray}
(\not{p}-m)\psi&=&0\,,
\end{eqnarray}
which will describe accordingly  the spin-cascade $(1/2^-,3/2^-)$,
\begin{eqnarray}
\left[(p^2-m^2)[g]^{\nu}_{\ \, \mu}-p_\mu p^\nu\right] \otimes
(\not{p}-m)\psi^\mu
&=&0, \nonumber\\
\psi^\mu&=&
A^\mu\otimes\psi\, .
\label{casc_eq}
\end{eqnarray}
The inverse of the latter equation has the advantage to provide a
 propagator,
\begin{equation}
\Pi_{(1/2,3/2)}(p)
= \frac{1}{2m}
\frac{\left( -g_{\mu\nu}+\frac{1}{m^2}p_\mu p_\nu\right)(\not p+m)
}{p^2 -m^2 + i\epsilon }\, ,
\label{csc_prp}
\end{equation}
which is consistent with the Proca and Dirac projectors in
Eqs.~(\ref{RS_Proca}) and (\ref{RS_Dirac}), respectively.

Upon gauging, Eq.~(\ref{Proca_main}) becomes
\begin{equation}
\label{acoplada} \left[(\pi^2-m^2)[g]^{\nu}_{\ \, \mu}-
\pi_\mu\pi^\nu\right](\not{\pi}-m)\psi^\mu=0\,,
\end{equation}
where we dropped the $\otimes$ sign for simplicity.
If now one calculates  the characteristic determinant
$|\Gamma_{\nu\mu}|$ of the matrix that has as elements all the
highest derivative terms with the derivatives being replaced
by  $n_\mu$, one finds it to be zero.
This means that the latter equation has constraints built in
which prevent it from being a  genuine system of differential equations.
This deficit is removed upon finding the gauged auxiliary
conditions and substituting them back into the leading equation.

To do so we contract Eq.~(\ref{acoplada}) by $\pi_\nu$ with the result,
\begin{eqnarray}
\left[\pi_\mu(\pi^2-m^2)-(\pi_\mu\pi_\nu+ieF_{\nu\mu})\pi^\nu\right]
(\not{\pi}-m)\psi^\mu&=&0\nonumber \\
\label{auxiliar}
\left[m^2\pi_\mu+ieF_{\nu\mu}\pi^\nu\right](\not{\pi}-m)\psi^\mu&=&0\,
, \\
m^2\pi_\mu(\not{\pi}-m)\psi^\mu&=&
ieF_{\mu\lambda}\pi^\lambda(\not{\pi}-m)\psi^\mu\, . \nonumber
\end{eqnarray}
In order to incorporate  the gauged auxiliary condition
in  Eqs.~(\ref{auxiliar}) into the wave equation
we commute the canonical momenta to obtain
 $\pi_\mu\pi^\nu$ in Eq.~(\ref{acoplada}),
\begin{eqnarray}
\left[(\pi^2-m^2)[g]^{\nu}_{\ \,  \mu}-(\pi^\nu\pi_\mu+ie F_\mu^{\
\, \nu})\right](\not{\pi}-m)\psi^\mu&=&0\, ,
\nonumber\\
\left\{(\pi^2-m^2)[g]^{\nu}_{\ \, \mu}(\not{\pi}-m)-
\pi^\nu\pi_\mu(\not{\pi}-m)+ieF^{\nu}_{\ \, \mu}
(\not{\pi}-m)\right\}\psi^\mu&=&0\, ,
\nonumber\\
\left\{(\pi^2-m^2)[g]^{\nu}_{\ \, \mu}-\frac{ie}{m^2}\pi^\nu
F_{\mu\lambda}\pi^\lambda+ieF^{\nu}_{\ \, \mu}\right\}
(\not{\pi}-m)\psi^\mu&=&0\, .
\end{eqnarray}
The following terms  contribute to the characteristic determinant
 $|\Gamma^{\nu}_{\ \, \mu}|$ :
\begin{equation}
\pi^2\not{\pi}[g]^{\nu}_{\ \, \mu}+\frac{ie}{m^2}\pi^\nu
F_{\lambda\mu}\pi^\lambda\not{\pi} \ \to \ \Gamma^{\nu}_{\ \,
\mu}=n^2\not{n}[g]^{\nu}_{\ \, \mu}+n^\nu\frac{ie}{m^2}
F_{\lambda\mu} n^\lambda\not{n}\, .
\end{equation}
The determinant to be calculated can be cast into the form
\begin{equation}
|\Gamma^{\nu}_{\ \, \mu}|=\left|n^2\not{n}[g]^{\nu}_{\ \, \mu}
+n^\nu\frac{ie}{m^2} F_{\lambda\mu}
n^\lambda\not{n}\right|=\left|\left(n^2[g]^{\nu}_{\ \, \mu}+
n^\nu\frac{ie}{m^2} F_{\lambda\mu}
n^\lambda\right)\otimes\left(\not{n}\right)\right|\,.
\end{equation}
The latter equation is no more but the direct product of
$[\Gamma^{\nu}_{\ \, \mu}]_{\mbox{\footnotesize Proca}}$ with
$[\Gamma^{\nu}_{\ \, \mu}]_{\mbox{\footnotesize Dirac}}$.\\
We now can use the general theorem on the determinant of the
direct (Kronecker) product of two matrices, here denoted by  $A$,
and $B$, of dimensionality $n$ and $q$, respectively,
 which tells one that
\begin{equation}
 |A\otimes B|=|A|^q|B|^n\, .
\label{Cron_prdt}
\end{equation}
In applying this formula to the direct product of
\begin{eqnarray}
\lbrack \Gamma^{\nu}_{\ \, \mu}\rbrack_{\mbox{\footnotesize
Proca}}=(n^2)^4\, , \quad &\mbox{\footnotesize and }&\quad \lbrack
\Gamma^{\nu}_{\ \, \mu}\rbrack_{\mbox{\footnotesize Dirac}}
=(n^2)^2\, ,
\end{eqnarray}
under consideration, we here find
\begin{equation}
[\Gamma^{\nu}_{\ \, \mu}]_{ \mbox{\footnotesize
Proca}\otimes\mbox{\footnotesize Dirac}}=(n^2)^{24}\, ,
\end{equation}
and arrive at the satisfactory result on its unconditional hyperbolicity.

This result can be independently confirmed by using appropriate
routines in  symbolic mathematical codes such as Maple in the
calculation of the $|\Gamma^\nu_{\,\, \mu}|$ determinant as
\begin{equation}
|\Gamma^{\nu}_{\ \, \mu}|=|n^2\not{n}[g]^{\nu}_{\ \,
\mu}+n^\nu\frac{ie}{m^2} F_{\lambda\mu}
n^\lambda\not{n}|=(n^2)^{24}\, .
\end{equation}

The conclusion is that spin-cascades are allowed to propagate causally
within the electromagnetic environment.
Same is valid for the Kronecker products of arbitrary
(finite) number of Proca equations on the basis of Eq.~(\ref{Cron_prdt}).
An interesting case is the one of two Proca equations where one
encounter the spin-cascade $(0,1,2)$, a case that may be of interest
to gravity.\cite{ADK}
Also this cascade will propagate causally within the electromagnetic
background, something which is out of reach for the
$\psi_{\mu_1\mu_2}$  sector characterized
by a single spin-2 at rest, a result due to.\cite{VZ2}

\section{Summary}
To summarize, we here found that Kronecker products of a finite number of
Proca equations with or without the Dirac equation allow for causal propagation
of the wave fronts of the associated solutions.
Such solutions, in carrying factorized Lorentz and spinor indices
are naturally coupled to the pion-nucleon, or photon-nucleon systems
and describe inseparable spin-cascades.
The corresponding  wave equations provide propagators that
are consistent with the projectors onto the states and
we expect this feature to facilitate the quantization procedure.
Spin cascades such like (1/2,3/2) and (0,1,2)
may be of interest both to particle spectroscopy and gravity.

\section{Acknowledgments}
We appreciate insightful  discussions with Mauro Napsuciale
on causality in the light of the Courant-Hilbert criterion.

Work supported by Consejo Nacional de Ciencia y
Tecnolog\'ia (CONACyT, Mexico) under grant number
C01-39820.

\nonumsection{References}


\begin{thebibliography}{000}

\bibitem{PF}M.\ Fierz, W.\ Pauli, Proc.\ Roy.\ Soc.\ (London)
\textbf{A173}, 211 (1939).

\bibitem{RS}W.\ Rarita, J.\ Schwinger, Phys.\ Rev.\ \textbf{60}, 61 (1941).

\bibitem{MC}  P.\ A.\ Moldauer, K.M.\ Case, Phys Rev. \textbf{102}, 279 (1956).

\bibitem{VZ1} G.\ Velo, D.\ Zwanziger,
Phys.\ Rev.\ \textbf{186}, 1337 (1969).


\bibitem{Weda}
J.\ Weda, \textit{Spin-3/2 particles and consistent $\pi
N\Delta$ and $\gamma N\Delta$-couplings\/}, Ph. D. thesis,
KVI, University of Croningen, July, 1999.

\bibitem{SJ}
K.\ Johnson, E.\ C.\ Sudarshan,
Annals of Physics \ \textbf{13}, 126 (1961).

\bibitem{MK} M.\ Kirchbach, M.\ Moshinsky, Yu.\ F.\ Smirnov,
Phys.\ Rev.\ {\bf D64 }, 114005 (2001).


\bibitem{KA} M.\ Kirchbach, D. V. Ahluwalia,
Phys.\ Lett. \ {\bf B529}, 124 (2002).


\bibitem{ADK}
D.\ V.\ Ahluwalia, N.\ Dadhich, M.\ Kirchbach, Int.\ J.\ Mod.\
Phys.\ D {\bf 11}, 1621 (2002).

\bibitem{SIK}  S.\ I.\ Kruglov,
{\it Symmetry and Electromagnetic Interactions of
Fields with Multi-Spin \/}
(Nova Science Publishers, Huntington, N.Y. 2001) p. 216.

\bibitem{CH} R.\ Courant, D.\  Hilbert,
{\it Methods of Mathematical Physics \/}
(Wiley Inter-science, New York, 1962).


\bibitem{RaSier} A.\ F.\ Ra$\tilde n$ada, G.\ Sierra,
Phys.\ Rev.\  {\bf D22}, 2416 (1980).

\bibitem{Kruglov} S.\ I.\ Kruglov, Int.\ J.\ Mod.\ Phys.\  {\bf A21},
1143 (2006).

\bibitem{Hurley_D4} W.\ J.\ Hurley, Phys.\ Rev.\ {\textbf D4}, 3605 (1971).

\bibitem{AE}D.\ V.\ Ahluwalia, D.\  J.\  Ernst,
Int.\ J.\ Mod.\ Phys.\  {\bf E2}, 397 (1996).



\bibitem{Ryder} L.\ H.\ Ryder, {\it Quantum Field Theory\/}
(Cambridge University Press, 1985)

\bibitem{hh} D.\ V.\ Ahluwalia, M.\ Kirchbach, Mod.\ Phys.\ Lett.\
{\bf A16}, 1377 (2001).

\bibitem{Mauro} M.\ Napsuciale, C.\ A.\ Vaquera-Araujo,
{\it Equations of motion as projectors and the gyromagnetic factor
g(s)=1/s from first principles \/},
hep-ph/0310106.


\bibitem {VZ2}G.\ Velo, D.\ Zwanziger,
Phys. Rev. \textbf{188}, 2218 (1969).


\end{thebibliography}
\end{document}